\documentclass[a4paper,11pt]{article}
\usepackage{graphicx,amssymb,amsmath,bm,latexsym,color,epsf,cite,comment}
\newcommand{\bea}{\begin{eqnarray}}
\newcommand{\eea}{\end{eqnarray}}
\newcommand{\vs}[1]{\vspace{#1 mm}}

\renewcommand{\b}{\beta}
\renewcommand{\c}{\gamma}
\renewcommand{\d}{\delta}

\newcommand{\ve}{\varepsilon}
\newcommand{\s}{\sigma}

\newcommand{\G}{\Gamma}

\newcommand{\la}{\lambda}
\newcommand{\pa}{\partial}
\newcommand{\nn}{\nonumber\\}

\newcommand{\lan}{\langle}
\newcommand{\ran}{\rangle}

\begin{document}

\begin{flushright}
NITEP 252
\end{flushright}
\begin{center}
{\Large\bf Standard Running, ``Physical Running'', \\
\vs{2}
Cosmological Constant and Newton Coupling}
\vs{10}

{\large
Hikaru Kawai$^{a,b,}$\footnote{e-mail address: hikarukawai@phys.ntu.edu.tw}
and
Nobuyoshi Ohta$^{c,}$\footnote{e-mail address: ohtan@ncu.edu.tw}
} \\
\vs{5}

$^a${\em Department of Physics and Center for Theoretical Physics, National Taiwan University, Taipei 106, Taiwan}

$^b${\em Physics Division, National Center for Theoretical Sciences, Taipei 106, Taiwan}
\vs{2}

$^c${\em Nambu Yoichiro Institute of Theoretical and Experimental Physics (NITEP), \\
Osaka Metropolitan University, Osaka 558-8585, Japan}

\vs{5}
%%%%%%%%%%%%%%%%%%%%%%%%%%%%%%%%
{\bf Abstract}
\end{center}

Recently it is asserted that the standard beta function does not describe the correct running
of the coupling constant in some theories. We show that the problem arises from the assumption $\mu=p$
($\mu$ is a renormalization point) and that a suitable choice of $\mu$ gives the correct running.
It is  also claimed that neither the cosmological constant nor Newton coupling run.
We argue that running can be discussed when we consider the curved spacetime.

\vs{10}

\setcounter{footnote}{0}

%%%%%%%%%%%%%%%%%%%%%%%%%%%%
%\section{Introduction}
%%%%%%%%%%%%%%%%%%%%%%%%%%%%

\section{Introduction}

The renormalization group (RG) is one of the most powerful tools in quantum field theory.
In the particle physics, the RG was introduced by Gell-Mann and Low~\cite{GL} as a convenient way
to absorb by a redefinition of the coupling constants the large logarithms of kinematical variables in loop
corrections to the scattering amplitudes. By this ``RG improvement,'' the substitution of the running coupling
into a tree-level expression gives the correct result with the subsequent identification of the RG scale
with an appropriate physical scale of the process.
Thus the couplings may be identified by a measurement of a physical amplitude.

In another approach by Wilson~\cite{Wilson}, the RG describes how the parameters in the effective action depend
on a UV cutoff and the way in which it regularizes divergences.
Quite often, the cutoff has a direct physical interpretation, such as the mass of the states integrated
out in the effective field theories.

In the ordinary theories, the scale dependences in these two approaches to RG are equivalent
in the high energy limit. This has been confirmed in various processes.
At energies comparable to the masses that are integrated out, the beta functions extracted from the Wilsonian RG
include threshold effects which encode the automatic decoupling of massive modes from the flow at scales
below the mass.

To regularize UV divergences, the dimensional regularization is most convenient because it preserves
symmetry properties of the theory. In this process, an unphysical energy scale $\mu$ is introduced.
Physics must be independent of this scale, but the renormalized amplitudes explicitly depend on it.
Consequently renormalized couplings must run with the scale to obtain physically meaningful results.
In principle, any choice of $\mu$ should be physically correct, but there is a convenient choice of $\mu$
such that the large logarithms are absent, giving the proper RG improvement. Sometimes, this point may be obscure.
There is another subtlety in using dimensional regularization to investigate dimensionful couplings.
This is because the minimal subtraction is a mass-independent renormalization scheme.
Therefore, when dealing with mass-dependent renormalization groups, other regularizations are often used
instead of dimensional regularization.
For example, we can calculate the quantum corrections in terms of
a dimensionally continued propertime or $\zeta$ function regularization.
The latter may be related to heat-kernel method.
More detailed discussions on the subtlety can be found in Ref.~\cite{BEGPPRSV}.

There has been some confusion related to these subtleties as described below.
In this paper, we would like to clarify how these problems could be resolved.

\section{``Physical Running'' of the coupling constants}

In a series of recent papers~\cite{BDP,BDMP1,BDMP2}, it is asserted that the standard beta function
does not give the correct running coupling constant but one should use other ``physical running'' coupling.
This gives  the impression that the standard beta function does not give the ``correct'' effective
coupling that actually describes physical quantities. Here we point out that this is due to the assumption
$\mu=p$ ($\mu$ is a renormalization point) even when other choice is suitable.

The point is best explained by the example of amplitude given in \cite{BDMP2}. Consider the amplitude
\bea
\label{amp1}
{\cal M}(p) = \la(\mu)+a \la^2(\mu)\log\left(\frac{m^2}{\mu^2}\right)
+b \la^2(\mu)\log\left(\frac{p^2}{\mu^2}\right)
+c \la^2(\mu)\log\left(\frac{p^2}{m^2}\right),
\eea
where $\mu$ comes from dimensional regularization and $m$ is either a mass that is present in the theory
or an IR regulator. Such an amplitude could arise, for example, from the application of the $\overline{\rm MS}$ scheme.
According to the spirit of the renormalization group (RG), the amplitude should not depend on $\mu$.
This leads to the condition
\bea
\mu\frac{d\la}{d\mu}- 2(a+b) \la^2 + 2 \mu\frac{d\la}{d\mu} \left[
a \la\log\left(\frac{p^2}{\mu^2}\right)
+b \la\log\left(\frac{p^2}{\mu^2}\right)
+c \la\log\left(\frac{p^2}{m^2}\right)\right]=0.
\eea
The last term proportional to the $\mu\frac{d\la}{d\mu}$ is higher order in $\la$.
So we find the standard beta function up to the second order in $\la$:
\bea
\b_\la =\mu\frac{d}{d\mu}\la(\mu)= 2(a+b) \la^2.
\label{beta1}
\eea
It is claimed that the $\mu$-dependence of the coupling contains a spurious part (the one proportional
to $a$) that does not reflect a momentum dependence in the amplitude and misses the momentum dependence
of the term proportional to $c$.
In fact, in \cite{BDP,BDMP1,BDMP2} by choosing $\mu \approx p$ they obtain
\bea
{\cal M}(p) = \la(p)+(a-c) \la^2(p)\log\left(\frac{m^2}{p^2}\right),
\label{wb}
\eea
which contains a large logarithmic term for large $p$.
It is proposed that we should absorb the $m$-dependent logarithm in~\eqref{amp1} in the definition
of the renormalized coupling
\bea
\la'(\mu) = \la(\mu) +(a-c) \la^2(\mu) \log\left(\frac{m^2}{\mu^2}\right),
\label{coup}
\eea
and then the amplitude reads
\bea
{\cal M}(p) = \la'(\mu)+(b+c) \la'{}^2(\mu)\log\left(\frac{p^2}{\mu^2}\right).
\label{amp2}
\eea
The beta function for $\la'$ is derived from the $\mu$-independence of the amplitude as
\bea
\b_{\la'}=\mu\frac{d}{d\mu}\la'(\mu)= 2(b+c) \la'{}^2,
\label{beta2}
\eea
up to the second order in $\la'$.
We note that given the redefinition~\eqref{coup}, this beta function can be derived also from \eqref{beta1}:
\bea
\mu\frac{d}{d\mu}\la'(\mu)= \mu\frac{d}{d\mu}\la(\mu) - 2(a-c) \la'{}^2,
\label{beta3}
\eea
which, upon substituting Eq.~\eqref{beta1}, reproduces \eqref{beta2}.
So both beta functions are equivalent.
The authors claim that it is the beta function~\eqref{beta2} that correctly tracks the momentum-dependence
of the amplitude.

However it is actually the assumption $\mu \approx p$ that causes the problem in \eqref{wb}.
The spirit of the RG is that no physical quantity should depend on $\mu$, which can be chosen as we wish.
In this sense, restricting only to $\mu \approx p$ sometimes makes the trouble, and the beta function
in \eqref{beta1} has no problem.
One has to be careful about how we define the running coupling.

In the standard RG approach, we define the effective coupling by requiring the large logarithm
in the amplitude is absent. This can be achieved by choosing $\mu$ as
\bea
a \log\left(\frac{m^2}{\mu^2}\right) +b \log\left(\frac{p^2}{\mu^2}\right)
+c \log\left(\frac{p^2}{m^2}\right)=0,
\eea
which gives the solution
\bea
(a+b)\log\left(\frac{\mu^2}{m^2}\right) = (b+c) \log \left(\frac{p^2}{m^2}\right).
\label{sol1}
\eea
We emphasize again that we can choose any $\mu$ for our convenience, and in particular we can choose it
such that the large logarithm is absent. This is the spirit of the RG.
The solution of the RG equation~\eqref{beta1} is
\bea
\frac{1}{\la}=\frac{1}{\la_0} - (a+b)\log\left(\frac{\mu^2}{m^2}\right),
\label{sol2}
\eea
where $\la_0$ is an integration constant.
Substituting \eqref{sol1} into this, we get
\bea
\frac{1}{\la}=\frac{1}{\la_0} - (b+c) \log\left(\frac{p^2}{m^2}\right),
\label{sol3}
\eea
which gives the correct momentum dependence.

In the scheme using the redefined coupling~\eqref{coup}, the requirement of the absence of large logarithm
in the amplitude~\eqref{amp2} gives $\mu=p$. The solution to the renormalization equation~\eqref{beta2} is
\bea
\frac{1}{\la'} = \frac{1}{\la_0'} - (b+c) \log\left(\frac{\mu^2}{m^2}\right),
\eea
where $\la_0'$ is an integration constant.
Setting $\mu=p$, we get
\bea
\frac{1}{\la'} = \frac{1}{\la_0'} - (b+c) \log\left(\frac{p^2}{m^2}\right) ,
\eea
which completely agrees with the momentum dependence in the standard result~\eqref{sol3}.
%(The $m$-dependent term in \eqref{sol3} could be absorbed into $\la_0$.)

Thus it may appear that the modified beta function \eqref{beta2} directly gives the momentum dependence
of the coupling and the amplitude,
while the standard beta function \eqref{beta1} does not. The claim that the standard beta function does not
give the correct ``physical running'' arises from the assumption $\mu \approx p$.
It is more convenient to choose $\mu$ such that the loop correction becomes small.
The RG equation simply means that physical quantities should not depend on $\mu$.
So any choice of $\mu$ should give the same physical quantities; the truth is that there is
a convenient choice of $\mu$ such that the large logarithm is absent. The choice depends on the process
but there is no superiority between the standard RG and the modified one if one does not forget the spirit of RG,
which is the $\mu$-independence of physical quantities.

\section{Cosmological constant and Newton coupling}

In recent papers~\cite{D1,D2}, an interesting observation is made as to the scale dependence of
the vacuum energy and the Newton coupling.\footnote{The RG equation is used to study the quantum effects
of gravity in the context of the asymptotically safe gravity~\cite{Reuter,NR,perbook,rsbook}.}
Here we discuss the vacuum energy along the line of the previous section.
Using $g_{\mu\nu}=\eta_{\mu\nu}+h_{\mu\nu}$, the cosmological term (to be precise, vacuum energy term)
in the action can be expanded as
\bea
\sqrt{-g} \rho = \rho\left(1+\frac12 h_\s^\s +\frac18 (h_\s^\s)^2 -\frac14 h_{\s\la}h^{\s\la} + \ldots\right).
\eea
One can study the renormalization of the vacuum energy by examining the one-point
etc. diagram in the weak field limit. The simplest way to calculate the vacuum energy is to consider
a single $h_{\mu\nu}$ field coupled to a massive scalar field with mass $m$.
Using the three-point vertex coupling coupling two scalar fields to the graviton fluctuation,
the one loop corrections to the vacuum energy is found to be (see, for example, \cite{Martin})
\bea
\d\rho = -\frac{m^4}{32\pi^2}\left[ -\c +\log(4\pi)+\log\frac{\mu^2}{m^2}+\frac32 \right],
\label{one}
\eea
in the MS scheme, where we subtract only the pole part in the dimensional regularization,
and $\mu$ is the renormalization mass scale.
Though this result depends on the parameter $\mu$,
the physical vacuum energy $\rho+\d\rho$ does not depend on $\mu$. (See Eq.~\eqref{pr}.)
This quantity is the vacuum energy on the flat spacetime.
Therefore the only mass scale is $m$, and the one-loop correction~\eqref{one} can be made small by setting $\mu=m$.
Thus the vacuum energy does not run in this definition.

On the other hand, let us reconsider the scale dependence of the vacuum energy and the Newton coupling
in the context of the effective action for the curved spacetime. We consider the action
\bea
S=\int d^4 x \sqrt{-g} \left( \frac{1}{16\pi G} R - \rho + \mbox{matter fields} \right),
\eea
where $G$ is the Newton coupling.
The generating functional of the one-particle irreducible Green's functions is denoted as
\bea
\Gamma[g_{\mu\nu}, \phi],
\eea
where $\phi$ stands for matter fields.
The expectation value of the vacuum energy-momentum tensor is given by
\bea
\lan T_{\mu\nu}(x) \ran_g = - \left. \frac{2}{\sqrt{-g(x)}} \frac{\d \G}{\d g^{\mu\nu}(x)} \right|_{\phi=0},
\eea
where $g_{\mu\nu}(x)$ is not necessarily the one that makes $\G$ stationary.
The vacuum energy from the matter and cosmological term is defined by
\bea
\lan T_{\mu\nu}^{\mbox{\small matter}+\rho}(x) \ran_g = \lan T_{\mu\nu}(x) \ran_g +\frac{1}{8\pi G} G_{\mu\nu},
\eea
where $G_{\mu\nu}$ is the expectation value of the Einstein tensor.

Suppose that our background is a maximally symmetric space, and then we expect that
\bea
\lan T_{\mu\nu}^{\mbox{\small matter}+\rho} (x) \ran_g = - \ve g_{\mu\nu}(x),
\eea
where $\ve$ is a constant and may be regarded as the vacuum energy, as is clear if we consider
$\mu=\nu=0$ component.
Note that there are two contributions to the vacuum energy from the matter and the cosmological constant.
For general backgrounds, the spacetime average of the vacuum energy is given by
\bea
\ve[g] = -\frac{1}{4} \frac{1}{\int d^4 x \sqrt{-g}}
\int d^4 x \sqrt{-g}\, g^{\mu\nu} \lan T_{\mu\nu}^{\mbox{\small matter}+\rho}(x) \ran_g
\eea
Here $\ve[g]$ is a functional of spacetime metric, and if the spacetime has curvature radius roughly~$L$
($R\sim \frac{1}{L^2}$),
we can consider it as a function of $L$. In this sense, we can discuss the scale dependence of the vacuum energy.

To be more concrete, let us look at the divergences from the free scalar field of mass $m$. It takes the form
\bea
\ve_{\rm 1-loop} = a \Lambda^4 + \left(b m^2 +c \frac{1}{L^2} \right) \Lambda^2
+\left(d_1 m^4 +d_2 m^2 \frac{1}{L^2} +d_3 \frac{1}{L^4}\right) \log\frac{f(m^2,\frac{1}{L^2})}{\Lambda^2}
+ \ldots,
\eea
where $a, b, c, d_1, d_2, d_3$ are constants, $f(m^2,\frac{1}{L^2})$ is a homogeneous function
of degree 1 such as $f= c_1 m^2 +\frac{c_2}{L^2}$ though its explicit form depends on the shape of the spacetime,
and the ellipsis denote finite terms in the infinite limit of the cutoff $\Lambda$.
We can write
\bea
\log\frac{f(m^2,\frac{1}{L^2})}{\Lambda^2} = \log \frac{f(m^2,\frac{1}{L^2})}{\mu^2}+ \log\frac{\mu^2}{\Lambda^2},
\eea
where $\mu$ is a renormalization mass scale.
Now the term $d_3 \frac{1}{L^4}\log\frac{\mu^2}{\Lambda^2}$ is removed by the renormalization of the terms
$R^2, R_{\mu\nu}^2$,
the terms $(c \Lambda^2 + d_2 m^2 \log\frac{\mu^2}{\Lambda^2})\frac{1}{L^2}$ by the Newton coupling,
and $a \Lambda^4 +b m^2 \Lambda^2+ d_1 m^4 \log\frac{\mu^2}{\Lambda^2}$ by the vacuum energy.
As a result, the renormalized vacuum energy is given by
\bea
\ve_{\rm 1-loop}^{\rm ren} =\left(d_1 m^4+ d_2 m^2\frac{1}{L^2}+d_3 \frac{1}{L^4}\right)
\log \frac{f(m^2,\frac{1}{L^2})}{\mu^2} +h\left(m, \frac{1}{L}\right),
\label{1loop}
\eea
where $h(x,y)$ is a homogeneous function of degree 4.
Since the bare couplings are given as
\bea
&& \rho_{\rm bare} = \rho - a \Lambda^4 - b m^2 \Lambda^2 - d_1 m^4 \log\frac{\mu^2}{\Lambda^2}, \nn
&& \frac{1}{16\pi G_{\rm bare}} = \frac{1}{16\pi G}- c \Lambda^2 - d_2 m^2 \log\frac{\mu^2}{\Lambda^2},
%\a_0 &= \a- \star \log\frac{\mu^2}{\Lambda^2}, \nn
%m_0^2 &= m^2- \star G \Lambda^4 - \star G m^2 \Lambda^2 - \star G \log\frac{\mu^2}{\Lambda^2},
\eea
the RG equations take the form
\bea
&& 0=\mu\frac{\pa}{\pa\mu} \rho_{\rm bare} = \mu\frac{\pa}{\pa\mu}\rho - 2 d_1 m^4, \nn
&& 0=\mu\frac{\pa}{\pa\mu} \left(\frac{1}{16\pi G_{\rm bare}}\right)
 = \mu\frac{\pa}{\pa\mu}\left(\frac{1}{16\pi G}\right) - 2 d_2 m^2.
\label{rgfinal}
\eea
(Here we have written only the contributions of the scalar field.)

The solution to the RG equations are given as
\bea
\rho = \rho_0 + d_1 m^4 \log \frac{\mu^2}{m^2}, \qquad
\frac{1}{16\pi G} = \frac{1}{16\pi G_0} + d_2 m^2 \log\frac{\mu^2}{m^2},
\label{solrg}
\eea
where $\rho_0$ and $G_0$ are integration constants.
We may choose $\mu^2$ such that the 1-loop correction~\eqref{1loop} is small:
\bea
\mu^2 =f\left(m^2,\frac{1}{L^2}\right),
\eea
giving
\bea
\rho = \rho_0 + d_1 m^4 \log \frac{f(m^2,\frac{1}{L^2})}{m^2}, \qquad
\frac{1}{16\pi G} = \frac{1}{16\pi G_0} + d_2 m^2 \log\frac{f(m^2,\frac{1}{L^2})}{m^2}.
\label{pr}
\eea
Thus the vacuum energy and the Newton coupling run if we consider the curved spacetime.

There is another point to be noticed. As we have shown in our recent papers~\cite{KO1,KO2},
it does not make sense to discuss the RG flow for the vacuum energy and Newton coupling separately
because their values themselves change by the wave function renormalization. For example, we can keep the vacuum energy
constant along the RG trajectory by suitably choosing the wave function renormalization.
However in this case the Newton coupling does change along the RG flow. Only those quantities
independent of the wave function renormalization make physical sense.
That is $\eta \equiv 16\pi G\sqrt{\rho}$, which does run.

\section*{Acknowledgments}

We would like to thank Roberto Percacci for useful correspondence.
H.K. thanks Prof. Shin-Nan Yang and his family for their kind support through the Chin-Yu
chair professorship. H.K. is partially supported by JSPS (Grants-in-Aid for Scientific Research
Grants No. 20K03970), by the Ministry of Science and Technology, R.O.C.
(MOST 111-2811-M-002-016), and by National Taiwan University.
The work of N.O. was supported in part by the Grant-in-Aid for Scientific Research Fund of the JSPS (C) No. 20K03980.

\end{document}